\documentclass[twocolumn,amsmath,amssymb,prb]{revtex4}
\usepackage{dcolumn}
\usepackage{epsfig}
\usepackage{graphics}
\usepackage[latin1]{inputenc} 
\usepackage[T1]{fontenc}
\usepackage{bm}
\begin{document}

\title{Structural, vibrational and thermal properties of densified silicates : insights from Molecular Dynamics}
\author{M. Bauchy}
 \affiliation{Laboratoire de Physique Théorique de la Matière Condensée,
Université Pierre et Marie Curie,  Boîte 121, 4, Place Jussieu, 75252
Paris Cedex 05, France}
\date{\today}

\begin{abstract}
Structural, vibrational and thermal properties of densified sodium silicate (NS2) are investigated with classical molecular dynamics simulations of the glass and the liquid state. A systematic investigation of the glass structure with respect to density was performed. We observe a repolymerization of the network manifested by a transition from a tetrahedral to an octahedral silicon environment, the decrease of the amount of non-bridging oxygen atoms and the appearance of three-fold coordinated oxygen atoms (triclusters). Anomalous changes in the medium range order are observed, the first sharp diffraction peak showing a minimum of its full-width at half maximum according to density. The previously reported vibrational trends in densified glasses are observed, such as the shift of the Boson peak intensity to higher frequencies and the decrease of its intensity. Finally, we show that the thermal behavior of the liquid can be reproduced by the  Birch-Murnaghan equation of states, thus allowing us to compute the isothermal compressibility.
\end{abstract}

\maketitle

\section{Introduction}

In the field of oxides, silicate glasses and melts have received a huge attention for their important applications in materials science and geophysics, such as magma dynamics and properties. Pressure (or density) is obviously one of the most important thermodynamic variable for geochemical processes in the mantle and crust. Indeed, interesting macroscopic properties of silicate melts, such as viscosity or diffusion, show significant changes with pressure. \cite{mysen_effect_1990, poe_silicon_1997}

Many experimental studies on silicate glasses, the base material for various multi-components silicate systems, have suggested that those macroscopic properties were related to atomic-scale structural changes \cite{yarger_coordination_1995} such as angles \cite{sharma_relationship_1979, kanzaki_phase_1998} or coordination number \cite{kanzaki_phase_1998, lee_structure_2004, lee_order_2003}. Densified sodium silicate is a very interesting system to be investigated as it shows the effect of polymerization and depolymerization\cite{kanzaki_phase_1998, lee_order_2003}. Indeed, in the silica network, Si tetrahedrons are connected by bridging oxygen atoms (BOs). Sodium silicate is usually described as a base silica network  which is depolymerized by the sodium atoms. In this view, sodium cations break Si-BO-Si bonds and induce non-bringding oxygen atoms (NBOs). On the contrary, pressure tends to repolymerize the network by a global increase of coordination numbers.

Sodium silicate glass has already been extensively studied at ambient pressure using Molecular Dynamics (MD). The first reported MD simulation of sodium silicate glass in 1979 was based on a very small system (200 atoms) but it is remarkable to see that it presented a very reasonable structural description of the glass. Since this work, the used potentials have been continuously improved to get a better reproduction of experimental results. Progress in computing facilities progressively allowed to reach longer time scales, thus making it possible to study diffusion at lower temperature and rheological properties \cite{soules_sodium_1981, soules_rheological_1983}. The possibility to simulate larger systems has also permitted to put in evidence the existence of inhomogeneities and preferential diffusion pathways for sodium atoms\cite{melman_microstructural_1991, smith_computer_1995, huang_structural_1991, jund_channel_2001, horbach_structural_2001, sunyer_characterization_2002, bauchy_pockets_2011}. Simulations from Cormack and co-workers\cite{du_medium_2004, cormack_alkali_2002, yuan_si-o-si_2003} have shown a very good agreement with experimental results on structure. Vibrational \cite{j._high-frequency_1994, zotov_heat_2002, smith_computer_1995} and elastic \cite{pedone_elastic_2008} properties of the glass at ambient pressure have also been studied and successfully compared to experimental data. Using superomputers, large scale classical simulations have recently been performed \cite{adkins_large-scale_2011}, as well as ab initio Molecular Dynamics simulations \cite{tilocca_sodium_2010, angeli_insight_2011, tokuda_inhomogeneous_2011}. However, to our knowledge, no systematic study of the evolution of the system according to pressure has been performed so far.

We present here Molecular Dynamics simulation allowing a systematic description of the structural, vibrational and thermodynamics properties of densified glassy and liquid sodium silicate. We focus in one particular composition (NS2) and study the properties with increasing density. Results show that a transition from tetrahedral to octahedral silicon environment occurs and that the medium range order shows anomalous changes. Vibrational properties are also found to be very sensitive to pressure and we report some trends about the behavior of the Boson peak according to density. Eventually, an equation of state model is proposed, thus allowing the computation of the isothermal compressibility.

The article is organized as follows. In section II, we present the numerical model and methodology that has been used. In section III, we report structural, topological and vibrational results of the glass. In section IV, thermodynamics and structural results of the liquid state are presented. Finally, section V summarizes these results.

\section{Simulation details}

As just mentioned, (Na$_2$O)$_x$ - (SiO$_2$)$_{1-x}$ with x=0.30 system has been chosen (close to the so-called NS2 system with x=0.33). The simulated system is composed of N = 3000 atoms (700 Si, 1700 O and 600 Na), placed in a cubic box of various lengths L to study different densities (from 1.5 g/cm$^3$ to 5.4 g/cm$^3$). To do so, all the simulations were run in the canonical ensemble (NVT). The room temperature density \cite{narottam_handbook_1986} of 2.466 g/cm$^3$  is obtained with L=34.43$\mathring{\text{A}}$. A computed pressure P = -1.6 GPa is found in the glass at this density.

To take into account the oxidation state of atoms \cite{cormack_alkali_2002}, partial charges are used for the Coulomb interaction, while the short-range Buckingham potential is of the form :

\begin{equation}
 V_{ij}(r) = A_{ij}exp(-\frac{r}{\varrho_{ij}})-\frac{C_{ij}}{r^6}
\end{equation} where $A_{ij}$, $\varrho_{ij}$ and $C_{ij}$ are parameters which have been fitted by Teter \cite{cormack_alkali_2002}. Usually, the Buckingham potential can induce spurious effects at high temperature (as V(r) can go to negative infinity when $r$ is close to zero, which leads to a collapse of the interacting atoms\cite{guissani_numerical_1996}. As described in \cite{du_medium_2004}, a repulsive term $B_{ij}/r^{n_{ij}}$ was introduced at short distance in order for the potential energy and its derivative to be continuous at $r_0$ to avoid this issue.

This potential has been extensively used by Cormack et al. \cite{cormack_alkali_2002, du_medium_2004} and has revealed a very good description of the glass at room density for various compositions. The effect of pressure on such systems using classical Molecular Dynamics has been considered \cite{diefenbacher_molecular_1998} only at high temperature for the NS4 silicate by using a Born-Mayer interaction potential fairly similar to the one that is presently used. While, at ambient pressure, the use of a Coulomb interaction with fixed partial charges is supported by the ionic character of the interactions and the absence of charge transfer, one may wonder to what extent fixed charges can be still considered with increasing pressure. While we are not aware of any report of densified silicates, a recent \textit{ab initio} Molecular Dynamics study (in which electrons and charge transfer are explicitly computed) on an oxide network-forming glass under high pressure \cite{brazhkin_nature_2008, trachenko_first-principles_2008} has not shown any deformations of the electronic cloud that would be significant enough for ambient pressure pseudopotentials to be modified. The mentioned example \cite{brazhkin_nature_2008, trachenko_first-principles_2008}, the consistency of the presently reported results and the fact that the present potential was successfully used to reproduce a diffusion anomaly of O and Si atoms with increasing density \cite{bauchy_pockets_2011, bauchy_viscosity_2012}, also observed in pure silica\cite{shell_molecular_2002} or water \cite{errington_relationship_2001}, suggest that a certain degree of confidence can be expected.

Classical Molecular Dynamics simulations were performed using the DLPOLY package \cite{smith_short_2006}. The equations of motion were integrated with the Verlet-Leapfrog algorithm, using a timestep of 2.0 fs. Coulomb interactions were evaluated by the Ewald summation method with a cutoff of 12.0 $\mathring{\text{A}}$. The short-range interaction cutoff was chosen at 8.0 $\mathring{\text{A}}$. As mentioned, the simulations were run in the canonical ensemble (NVT) with a Berendsen thermostat.

For each density, the system was first equilibrated at 6000 K during $10^6$ steps (2ns). Each melt was then continuously cooled down to the selected temperature (from 300 K to 4000 K) using a cooling rate of 10 K/ps.

\section{Glass}
\subsection{Real space properties}

\subsubsection{Total radial correlation functions}

\begin{figure}
\begin{center}
\epsfig{figure=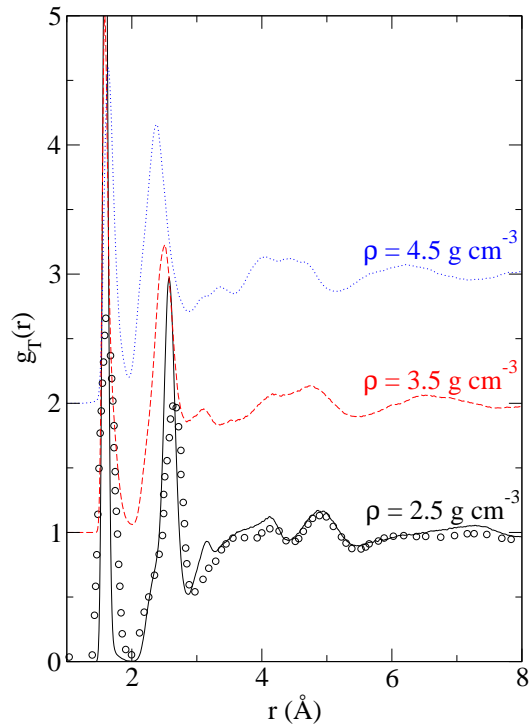,width=\linewidth}
\end{center}
\caption{\label{g} (Color online) Total radial correlation function of MD modeled sodium silicate glasses for increasing densities and comparison with neutron diffraction studies (white rounds) from the work of Wright et al. \cite{wright_neutron_1991} (Neutron diffraction data).}
\end{figure} 

The total correlation functions $g_{\text{T}}(r)$ for increasing densities are shown in Fig. \ref{g}. To check the validity of the simulated glass, comparison with experimental data (neutron diffraction from the work of Wright et al. \cite{wright_neutron_1991}) at room pressure was made. We recover the same level of agreement than in previous studies\cite{yuan_si-o-si_2003, du_medium_2004}. However, we notice an increased structured system with main peaks being sharper as compared to experiments. This comparison has also been done by Cormack \cite{du_medium_2004}. Using the same potential, a better agreement has been observed by broadening the total correlation functions.\cite{wright_experimental_1993} The position of the first Si-O peak is well reproduced, but is found to be sharper than in experiments. The position of the second O-O peak is also well reproduced, suggesting a realistic O-Si-O angle in simulation. On the other hand, simulation produces a peak at 3.1$\mathring{\text{A}}$ arising from Si-Si correlations (see below) which is not present in experiments but merged with other contributions in the region 3-4$\mathring{\text{A}}$. It means that the inter-tetrahedral angle Si-O-Si may be underestimated with respect to experiments. This angle turns out to be highly sensitive to the employed potential. A detailed discussion about the ability of the different potentials to reproduce the Si-O-Si angle can be found in \cite{yuan_si-o-si_2003}.

As density increases, the first Si-O peak does not show any shift in position but becomes broader, suggesting an increased disorder in the network, manifested by increased coordination numbers (integral of the first peak). As observed on the partial $g_i(r)$ distributions (see below), the second peak is shifted to lower r and becomes broader.

\subsubsection{Partial radial correlation functions}

\begin{figure*}
\begin{center}
\epsfig{figure=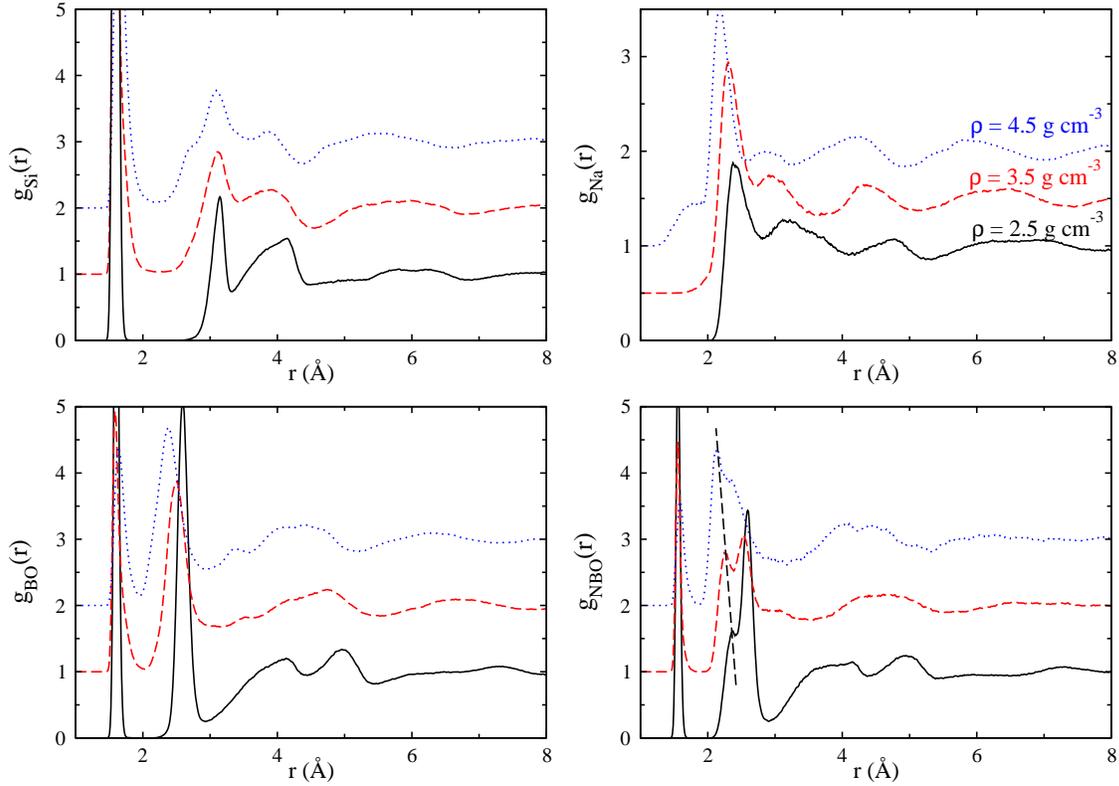, height=\linewidth, angle=-90}
\end{center}
\caption{\label{gi} (Color online) Partial radial correlation function $g_{\text{Si}}(r)$, $g_{\text{Na}}(r)$, $g_{\text{BO}}(r)$ and $g_{\text{NBO}}(r)$ at different selected densities $\varrho$ = 2.5, 3.5, 4.5 g/cm$3$. }
\end{figure*} 

The partial radial correlation functions g$_i$(r) have been computed from the pair correlation functions g$_{ij}$(r) :

\begin{equation}
 g_i(r) = \frac{1}{n} \sum_{j=1}^n g_{ij}
\end{equation} We have split the analysis according to BO and NBO. These functions are shown in Fig. \ref{gi} for increasing densities. While the position of the first peak in $g_{\text{Si}}$ (Si-O correlations) does not show any significant change, an increase in the shoulder on the lower r side of the second peak (Si-Si correlations) is observed as density increases, suggesting that the Si-O-Si angle decreases. As mentioned previously, the environment of the BO and NBO are studied separately using $g_{\text{BO}}$ and $g_{\text{NBO}}$. For both, the position of the first peak (O-Si correlations) remains the same, but important changes take place with density change for the second-neighbor correlation. The second peak (O-O correlations) is shifted to lower r and the distribution becomes broader. In the $g_{\text{NBO}}$ partial correlation function, one notices the growth of a peak (at 2.4$\mathring{\text{A}}$ for $\varrho$ = 3.5 g/cm$^3$) which contributes only to a shoulder of the main peak at 2.6$\mathring{\text{A}}$ for $\varrho$ = 2.5 g/cm$^3$. This also suggests that densification affects the O-Si-O and Si-O-Si angles rather than the Si-O distance between nearest neighbors. Finally, the first peak of $g_{\text{Na}}(r)$, associated with Na-O correlations, is shifted to lower r. The decrease in the Na-O distance with pressure has also been observed using NMR by Lee\cite{lee_structure_2004}. The Na-centered pair distribution functions are highly sensitive to density change and this is not surprising as it involves non-directional bonds. However, we notice that the increase of density leads to a better defined first peak whose height increases with the density.

\subsubsection{Coordination numbers}

\begin{figure}
\begin{center}
\epsfig{figure=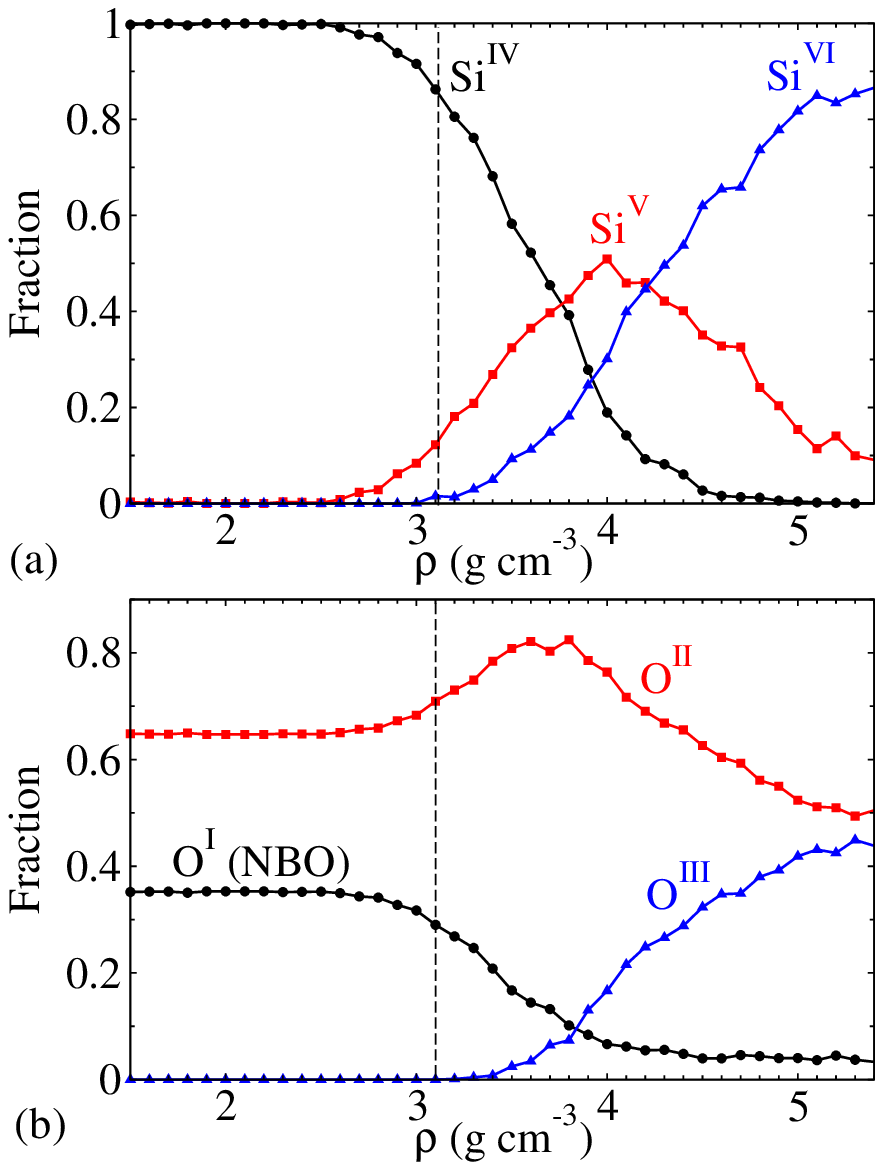,width=\linewidth}
\end{center}
\caption{\label{n} (Color online) Distribution of IV, V and VI-fold coordinated silicon atoms (a) and of I, II and III-fold coordinated oxygen atoms (b) with respect to density. Sodium atoms are not taken into account in the enumeration of the neighbors, so that O$^{\text{I}}$ refer to NBOs.}
\end{figure}

In pure silica, the network in fully connected and the coordination number CN of Si and O atoms are found to be 4 and 2, in agreement with the stoichiometry of the glass (CN$_{\text{Si}}$N$_{\text{Si}}$ = CN$_{\text{O}}$N$_{\text{O}}$). This is not the case in sodium silicates since Na atoms create NBOs, thus disrupting the network.

The distributions of IV, V and VI-fold coordinated silicon atoms (Si$^{\text{IV}}$, Si$^{\text{V}}$ and Si$^{\text{VI}}$) can be obtained by enumerating the number of oxygen neighbors in the first coordination shell of each silicon atom. These populations are shown in Fig. \ref{n}a for each CN. The fraction of tetrahedral Si$^{\text{IV}}$ atoms starts to drop from $\varrho$ = 2.7 g/cm$^{3}$ (P $\simeq$ 1 GPa). At the same density, the fraction of Si$^{\text{V}}$ atoms grows and reaches a maximum around $\varrho$ = 4.0 g/cm$^{3}$ (P $\simeq$ 28 GPa) prior to a continuous decrease as density increases. The fraction of octahedral Si$^{\text{VI}}$ atoms increases from $\varrho$ = 3.1 g/cm$^{3}$ (P = 5 GPa) and this basic structure becomes predominant at high density. These trends are rather usual in densified silicates. In amorphous silica, simulations from Tse\cite{tse_high-pressure_1992} predicted the increase of the Si CN to 5 at 15 GPa and up to 6 at 20 GPa. That trend was confirmed by simulations from Horbach\cite{horbach_molecular_2008}. The appearance of Si$^{\text{V}}$ and Si$^{\text{VI}}$ in densified sodium silicate has been confirmed experimentally using NMR. \cite{lee_order_2003, kanzaki_phase_1998}

The environment of oxygen atoms has been analyzed in the same fashion, i.e. by enumerating the number of silicon atoms in the first coordination shell of each oxygen atom. Here, Na atoms are not taken into account, this in order to distinguish BO from NBO and thus to split the Si CN analysis from the one involving the Q$^n$ speciation which will be detailed below. Thus, O$^{\text{I}}$ refers to the oxygen atoms that are surrounded by only one silicon atom (i.e. NBO atoms). At low density, the fraction of NBO can be determined by x, the amount of soda, as each sodium atom creates one NBO. The fraction of NBO f$_{\text{NBO}}$ is thus given by f$_{\text{NBO}}$ = N$_{\text{NBO}}$/N$_\text{O}$ = 2x/(2-x). At x = 0.3, f$_{\text{NBO}}$ $\approx$ 0.35, which is consistent with simulation results in Fig. \ref{n}b. The fraction of O$^{\text{I}}$ drops for densities larger than $\varrho$ = 2.6 g/cm$^{3}$. At this density, the fraction of O$^{\text{II}}$ starts to increase, reaches a maximum at $\varrho$ = 3.8 g/cm$^{3}$ and decreases at higher density. The present findings clearly indicate a repolymerization of the network through the creation of density induced Si-BO-Si connections at the expense of Si-NBO ones. They are consistent with the decrease of the fraction of NBO found experimentally from NMR in densified silicates \cite{lee_structure_2004}. As a consequence, the model of the network modifier Na atom simply given by stoichiometry (one Na atom involving the appearance of one NBO atom) does not remain valid for $\varrho >$ 3 g/cm$^{3}$. Ultimately, O$^{\text{III}}$ are found at high density and their fraction grows up to nearly 50$\%$ at $\varrho$ = 5.5 g/cm$^{3}$. Note that 3-fold O atoms (termed triclusters) have already been found both in experiments and in simulations, for example in aluminosilicate glasses \cite{tossell_o_2005}.

At $\varrho$ = 5.5 g/cm$^{3}$, the fraction of NBOs is very low ($\simeq$ 3$\%$) so that the Si/O network can be considered as being fully connected as in pure silica, thus allowing us to check the agreement between the stoichiometry of the system (SiO$_{2.43}$) and the computed coordination numbers. On average, we find CN$_{\text{Si}}$ = 5.90 and CN$_{\text{O}}$ = 2.43 so that the stoichiometry of the glass is satisfied (CN$_{\text{Si}}$N$_{\text{Si}}$ $\approx$ CN$_{\text{O}}$N$_{\text{O}}$).

These results show that the network undergoes strong topological changes as density increases. The initial tetrahedral silicon environment becomes octahedral at high density, consistently with the decrease of the O-Si-O angle (see below). On the other hand, a transition from 2-fold to 3-fold coordinated oxygen atoms is observed, which is once again consistent with the decrease of the Si-BO-Si angle (see below).

\subsubsection{Q$^n$ populations}

\begin{figure}
\begin{center}
\epsfig{figure=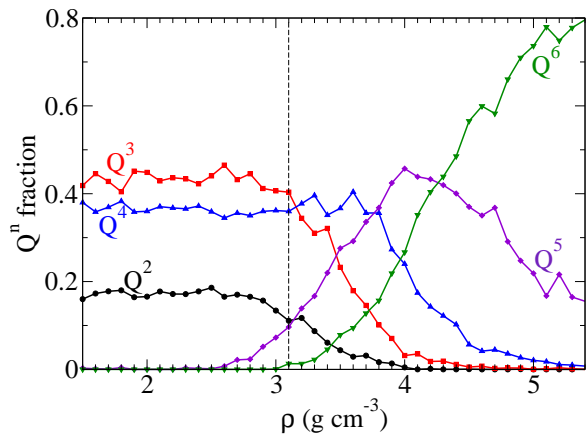,height=\linewidth, angle=-90}
\end{center}
\caption{\label{Qn} (Color online) Distribution of Q$^n$ populations with increasing density.}
\end{figure} 

As mentioned earlier, changes in the glass network can also be characterized by the Q$^n$ distribution analysis. We remind that a Q$^n$ silicon atom is defined as an atom linked with n bridging oxygen atoms. Defining BO and NBO at high density needs a careful analysis since 3-fold coordinated oxygen atoms can be found. NBOs are here thus defined as oxygen atoms connected to only one Si. BOs are defined as oxygen atoms that are not NBOs.

At ambient pressure, the Q$^n$ distribution usually range from a full Q$^4$ (the silica network) to Q$^0$ network, the orthosilicate glass, depending on the amount of soda x. At $\varrho$ = 2.5 g/cm$^{3}$, the Q$^0$, Q$^5$ and Q$^6$ populations were found to be negligible (less than $0.1\%$ in each case), Q$^{1,2,3,4}$ populations from simulation being given in Table \ref{Qn2} and compared with results from a previous simulation\cite{du_medium_2004}, with results of NMR studies\cite{maekawa_structural_1991} and with results \cite{du_medium_2004} from a random model proposed by Lacy \cite{lacy_statistical_1965}. First, we notice that our findings differ slightly with those obtained by Cormack\cite{du_medium_2004} using the same potential. The origin may be due to the fact that the system has a different thermal history (the cooling methodology is slightly different although the cooling rate is the same). However, both simulations are consistent with the random model. Differences with experimental data are important and these shifts have been found even in very large-scale simulations \cite{adkins_large-scale_2011}. They have been attributed to the fact that the high cooling rate used in simulations induces a structure with a higher effective temperature \cite{tan_effect_2004} so that the simulated Q$^n$ statistics is the one of a high temperature frozen liquid.

\begin{center}
\begin{table}[h]
\caption{\label{Qn2} Proportion of Q$^n$ populations at room density.}
\begin{tabular}{|l|l|l|l|l|}
\hline
Q$^n$ & Present MD & MD Cormack\cite{du_medium_2004} & NMR\cite{maekawa_structural_1991} & Random model\cite{lacy_statistical_1965}\\
\hline
Q$^1$ & 1.288 & 1.857 & 0.000 & 2.985\\
Q$^2$ & 18.598 & 15.571 & 4.776 & 16.716\\
Q$^3$ & 44.067 & 49.000 & 68.358 & 41.493\\
Q$^4$ & 35.908 & 33.571 & 26.567 & 37.910\\
\hline
\end{tabular}
\end{table}
\end{center}

When density changes, we observe that the changes in Q$^n$ populations are correlated with the change in O and Si coordination numbers (Fig. \ref{n}), i.e. they take place only for $\varrho >$ 3 g/cm$^{3}$. Q$^n$ populations do not show any significant changes at low density. At $\varrho$ = 2.5 g/cm$^{3}$, the Q$^5$ proportion starts to increase however and reaches a maximum at $\varrho$ = 4.0 g cm$^{3}$. Again we notice a clear correlation between the Q$^5$ population and the proportion of Si$^{\text{V}}$. The Q$^6$ proportion only starts to increase from $\varrho$ = 3.1 g /cm$^{3}$, thus showing a behavior similar to the one of the proportion of Si$^{\text{VI}}$.

\subsubsection{Bond-angle distributions}

\begin{figure}
\begin{center}
\epsfig{figure=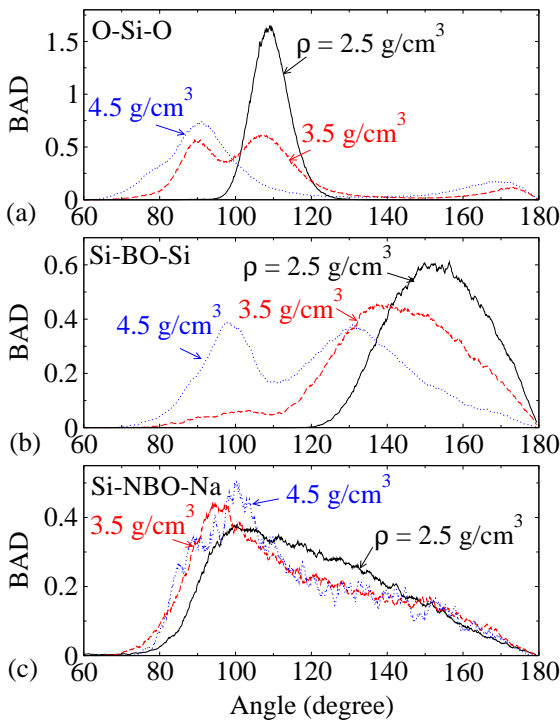,width=\linewidth}
\end{center}
\caption{\label{aijk} (Color online) O-Si-O (a), Si-BO-Si (b) and Si-NBO-Na (c) bond angle distributions at $\varrho$ = 2.5, 3.5 and 4.5 g cm$^{-3}$.}
\end{figure} 

\begin{figure}
\begin{center}
\epsfig{figure=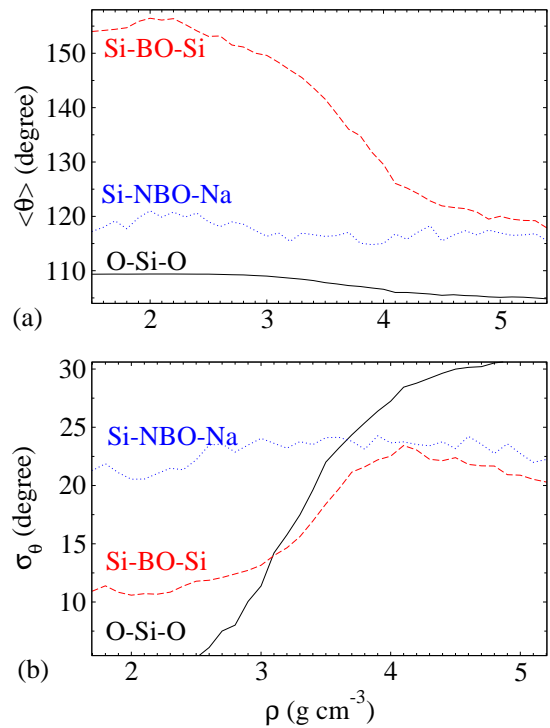,width=\linewidth}
\end{center}
\caption{\label{as} (Color online) First (a) and second (b) moment of the O-Si-O, Si-BO-Si and Si-NBO-Na bond angle distributions.}
\end{figure} 

We now focus on the bond-angle distributions (BAD) and their variations with density, which have been shown to be extremely sensitive in other tetrahedral systems \cite{mead_molecular_2006, tse_high-pressure_1992}. Even at ambient pressure, it allows to understand how the basic structures of the glass connect to each other. The O-Si-O BAD, which characterizes the Si tetrahedrons, is shown in Fig. \ref{aijk}a for three selected densities ($\varrho$ = 2.5, 3.5 and 4.5 g/cm$^{3}$). As expected at the lowest density ($\varrho$ = 2.5 g/cm$^{3}$), the distribution is sharp and the average O-Si-O angle is close to the ideal 109.5° tetrahedral angle (see also Fig. \ref{as}a). At intermediate densities however ($\varrho \approx$ 3.5 g/cm$^3$), the O-Si-O angle displays now a bimodal distribution with a peak still located at 109°, reminiscent of the initial tetrahedral structure, and a growing peak at 90°. The latter corresponds to the angle that is expected for an octahedral environment. An additional signature for this environment is provided by the contribution at 180° which appears for larger densities (Fig. \ref{aijk}a) and grows with $\varrho$. At high density ($\varrho$ = 4.5 g/cm$^{3}$, all silicon atoms display an octahedral environment with a single peak at 90° (and the vanishing of the tetrahedral peak at 109°) and the contribution at 180°. The second moment of the O-Si-O BAD is shown in Fig. \ref{as}b and grows from 5° at ordinary density ($\varrho$ = 2.5 g/cm$^{3}$) up to more than 30° at high density, suggesting the appearance of a pressure induced disorder manifested by an increased angular excursion around a mean value.

The Si-BO-Si angle characterizes the way two adjacent silicon tetrahedrons are connected. At ordinary density, the angle shows a broad distribution between 120° and 180° and centered at 153° (see also Fig. \ref{as}a), compared to the 142° experimental value from NMR \cite{pettifer_nmr_1988, farnan_quantification_1992}. This difference was also observed by previous simulations, as reviewed in \cite{yuan_si-o-si_2003} and is consistent with the over estimated value of the Si-Si distance. Like the O-Si-O angle, the Si-BO-Si angle displays a bimodal distribution as density increases. The BAD shows a second peak close to 100° at high density ($\varrho$ = 4.5 g/cm$^{3}$). This contribution is absent at intermediate densities ($\varrho$ = 3.5 g/cm$^{3}$, see Fig. \ref{aijk}b) but the trend with $\varrho$ is clearly correlated with the population of O$^{\text{III}}$ (see Fig. \ref{n}b and Fig. \ref{as}a). The decrease of the average value of the Si-BO-Si angle has also been observed experimentally in silicates \cite{stolper_nature_1987} and for GeO$_2$ \cite{sharma_relationship_1979}.

On the contrary, the Si-NBO-Na BAD does not show any significant change with density.

\subsubsection{Orientational parameter}

\begin{figure}
\begin{center}
\epsfig{figure=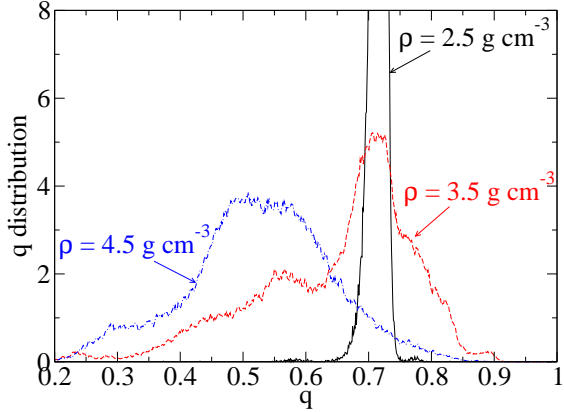,height=\linewidth, angle=-90}
\end{center}
\caption{\label{q} (Color online) q factor distributions for increasing densities.}
\end{figure}

\begin{figure}
\begin{center}
\epsfig{figure=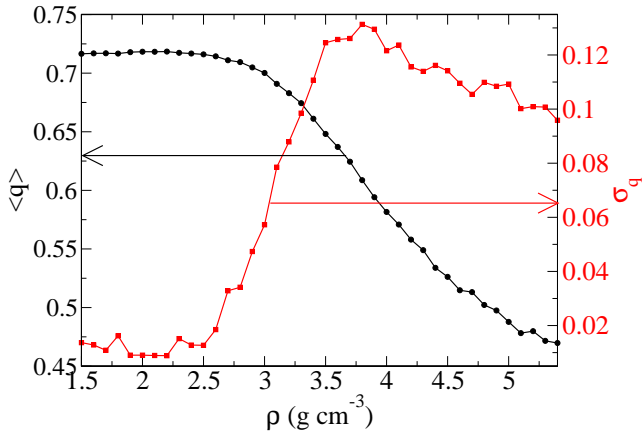,height=\linewidth, angle=-90}
\end{center}
\caption{\label{qs} (Color online) First (left axis) and second (rigt axis) moment of the q factor distributions for increasing densities.}
\end{figure} 

An interesting means to analyze the tetrahedral to octahedral conversion in liquids and glasses is provided by the orientational order parameter $q$ (introduced by Chau and Hardwick \cite{chau_new_1998} and rescaled in \cite{errington_relationship_2001}), which quantifies the extent to which a molecule and its four nearest neighbors adopt a tetrahedral arrangement. It is defined by :

\begin{equation}
 q = 1 - \langle \frac{3}{8} \sum \limits_{i=1}^3 \sum \limits_{k=j+1}^4 ( \cos \theta_{ijk} + \frac{1}{3} ) \rangle
\end{equation} where $\theta_{ijk}$ is the angle formed by the central Si atom $i$ and its oxygen nearest neighbors $j$ and $k$, the brackets representing an average over the central Si atoms i and over the time. This parameter is normalized so that its average varies between 0 (randomly arranged bonds) and 1 (perfect tetrahedral network). It has been used for the analysis of diffusivity anomalies in relationship with tetrahedral to octahedral changes in liquid water\cite{errington_relationship_2001}, silica\cite{shell_molecular_2002} and germania\cite{jabes_tetrahedral_2010}.

Fig. \ref{q} shows the calculated distribution of $q$ values for 3 selected densities ($\varrho$ = 2.5, 3.5 and 4.5 g/cm$^{3}$). At room density, the distribution exhibits only a sharp peak close to $q = 0.7$ (see Fig. \ref{qs}), which corresponds to near-perfect tetrahedral order, as also found for silica at ambient pressure\cite{shell_molecular_2002}. At $\varrho$ = 3.5 g/cm$^{3}$, the observed bimodal distribution suggests the existence of both tetrahedral Si (high $q$ peak) and higher coordinated Si (low $q$ peak with q $<$ 0.6). At $\varrho$ = 4.5 g/cm$^{3}$, the tetrahedral contribution has vanished. The second moment of the distributions according to density is shown in Fig. \ref{qs}. It characterizes the orientational disorder around central Si atoms, with $\sigma_{\text{q}}$ going from 0.02 at low density up to 0.1 at $\varrho$ = 4.5 g/cm$^{3}$, suggesting once again the appearance of a pressure induced disorder.

\subsection{Reciprocal space properties}

To investigate the structure of the glass on intermediate length scales, the neutron structure factor has been computed. The partial structure factors have been first calculated from the pair distribution functions $g_{ij}(r)$ :

\begin{equation}
S_{ij}(Q) = 1 + \varrho_0 \int_{0}^R 4\pi r^2 (g_{ij}(r)-1) \frac{\sin (Qr)}{Qr} F_{\text{L}}(r)\, \mathrm dr
\end{equation} where $Q$ is the scattering vector, $\varrho_0$ is the average atom number density and $R$ is the maximum value of the integration in real space (here $R = 15 \mathring{\text{A}}$). The $F_{\text{L}}(r) = \sin (\pi r / R) / (\pi r / R)$ term is a Lortch-type window function used to reduce the effect of the finite cutoff of $r$ in the integration\cite{wright_neutron_1988}. As discussed in \cite{du_compositional_2006}, the use of this function reduces the ripples at low $Q$ but induces a broadening of the structure factor peaks. The total neutron structure factor can then be evaluated from the partial structure factors following :

\begin{equation}
S_N(Q) = (\sum_{i,j=1}^n c_ic_jb_ib_j)^{-1} \sum_{i,j=1}^n c_ic_jb_ib_j S_{ij}(Q)
\end{equation} where $c_i$ is the fraction of $i$ atoms (Si, O or Na) and $b_i$ is the neutron scattering length of the species (given by 5.803, 4.1491 and 3.63 fm for oxygen, silicon and sodium atoms respectively\cite{sears_neutron_1992}).

\subsubsection{Neutron structure factor}

\begin{figure}
\begin{center}
\epsfig{figure=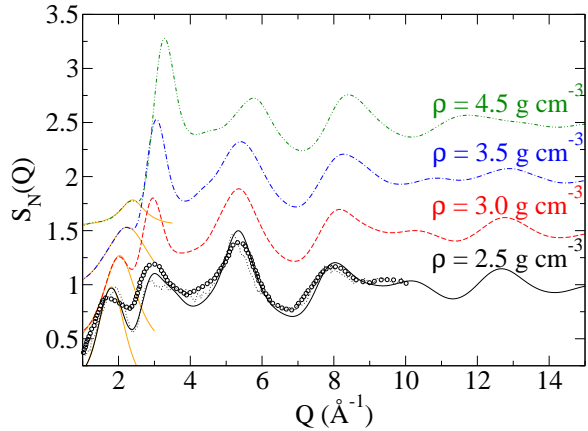,height=\linewidth, angle=-90}
\end{center}
\caption{\label{S} (Color online) Neutron structure factors for increasing densities. Neutron diffraction results from Wright et al. \cite{wright_neutron_1991} (open circles) and simulation results from Horbach at al. \cite{horbach_structural_2001} (dotted line) are shown for comparison. Examples of Lorentzian fits of the FSDP are displayed in orange.}
\end{figure} 

The total neutron structure factor $S_{\text{N}}$ for different increasing densities are shown on Fig. \ref{S}. The room density structure factor is compared both with the neutron diffraction results from Wright et al. \cite{wright_neutron_1991} and the simulated NS2 glass from Horbach et al. \cite{horbach_structural_2001}, using an alternative (BKS) potential. We note that the agreement between simulation and experiment is good. The agreement of the first peak with experiment is discussed in details in the FSDP section below. The second peak position is well reproduced (3.0$\mathring{\text{A}}$$^{-1}$ experimentally, compared to 3.0$\mathring{\text{A}}$$^{-1}$ from the present potential and 2.9$\mathring{\text{A}}$$^{-1}$ from the BKS potential). The third peak position is also very well reproduced (5.4$\mathring{\text{A}}$$^{-1}$ experimentally, compared to 5.3$\mathring{\text{A}}$$^{-1}$ from the present potential and 5.2$\mathring{\text{A}}$$^{-1}$ from the BKS potential).

As observed on Fig. \ref{S}, the density mainly influences the low wave vector part of the structure factors, which suggests that the main effects of density do not apply at short length scales. The shape in the high $Q$ limit (at $Q$>10$\mathring{\text{A}}$$^{-1}$ is nearly unchanged for $\varrho \leqslant$ 3.5 g/cm$^{3}$. All peaks are shifted to higher wave vector (lower r) as density increases, which is linked to the compaction of the network. Interestingly, the second moments of the different peaks do not show the same behavior with density. The so-called first sharp diffraction peak (FSDP) at very low $Q$ becomes broader and less intense with increasing density, as discussed below. The main peak around 3$\mathring{\text{A}}$$^{-1}$ becomes narrower as density increases, whereas the third one (around 5$\mathring{\text{A}}$$^{-1}$) becomes broader. These trends can be analyzed in more details from the partial structure factors (see below). The shape of the other peaks are almost unaffected by density.

\subsubsection{Partial structure factors}

\begin{figure}
\begin{center}
\epsfig{figure=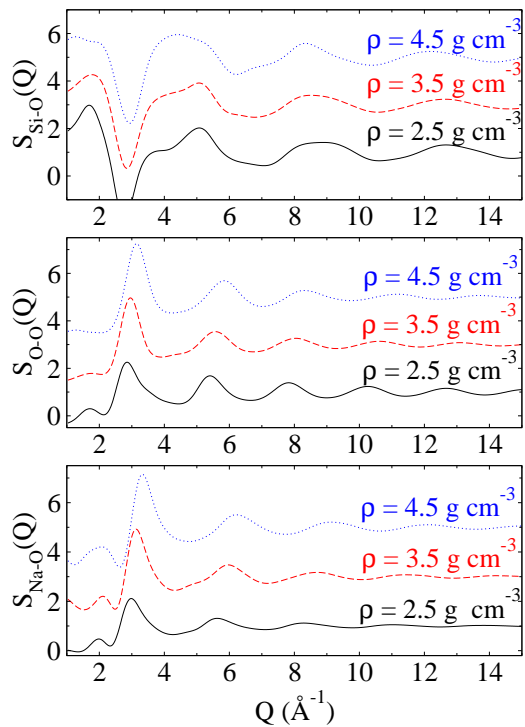,width=\linewidth}
\end{center}
\caption{\label{Sij} (Color online) Partial structure factors Si-O, O-O and Na-O for increasing densities.}
\end{figure}

Fig. \ref{Sij} shows the decomposition of the total structure factor into contributions of different pair structure factors $S_{\text{Si-O}}(Q)$, $S_{\text{O-O}}(Q)$ and $S_{\text{Na-O}}(Q)$ for different increasing densities. The partial $S_{\text{Si-Si}}(Q)$, $S_{\text{Na-Na}}(Q)$ and $S_{\text{Si-Na}}(Q)$ decay the fastest and have thus not been displayed. At normal density, the shape of these pair structure factors and the positions of the peaks are in excellent agreement with previously reported results from MD simulations \cite{du_first_2005}. At room pressure, the partial $S_{\text{Si-O}}(Q)$ shows the most significant variations with $Q$ both at short wave vector, correlated to the medium-range order of the silicate network, and at long $Q$, correlated to the strong short-range Si-O order. As already observed in the total structure factor, most of the peaks are also shifted to higher $Q$ (lower $r$) as density increases.

The decomposition of the total structure factor can serve to understand the behavior of the main peak ($\simeq 3\mathring{\text{A}}$$^{-1}$) and of the second main peak ($\simeq 5\mathring{\text{A}}$$^{-1}$) with density. Indeed, the peak at 3$\mathring{\text{A}}$$^{-1}$ in O-O and Na-O partial structure factors becomes narrower as density increases, an effect which is related to the increased structural medium-range order at high density. These peaks contribute the most of the second peaks of the total structure factor. On the other hand, the main contribution for the peak at 5$\mathring{\text{A}}$$^{-1}$ of the total structure factor arises from the second Si-O partial structure factor peak, which becomes broader as density increases. This apparent disorder may be attributed to the appearance of coexisting tetrahedral and octahedral Si-O environment as density increases.

\subsubsection{First sharp diffraction peak}

FSDPs are not simply the first of the many peaks of any diffraction pattern but display many anomalous behavior as a function of temperature, pressure and composition.\cite{s.r._extended-range_1995} Since the position of the FSDP $Q_{\text{FSDP}}$ is smaller than $Q_{\text{P}}$ (the position of the principal peak of the structure factor, associated to the nearest-neighbor distance), the FSDP corresponds to structural correlations on a larger length scale. This feature has been observed both in covalent\cite{elliott_origin_1991, sokolov_medium-range_1992} and ionic\cite{wilson_prepeaks_1994} amorphous system. In ionic systems, this medium range order has been associated to the forced separation between cations because of their mutual Coulomb repulsion, thus producing a prepeak in the cation-cation structure factor\cite{tatlipinar_atomic_1992}. Prepeaks can also arise from size effects of the atoms of the network \cite{iyetomi_atomic-size_1993}. However, the network formation itself can have a major role since the FSDP is also observed in the monoatomic tetravalent systems a-Si and a-Ge \cite{dixmier_hole_1992, uhlherr_extended-range_1994}. The FSDP origin is now usually explained by using a void-based model\cite{s.r._extended-range_1995, zaug_pressure-dependent_2008} in which ordering of interstitial voids occurs in the structure.

\begin{figure}
\begin{center}
\epsfig{figure=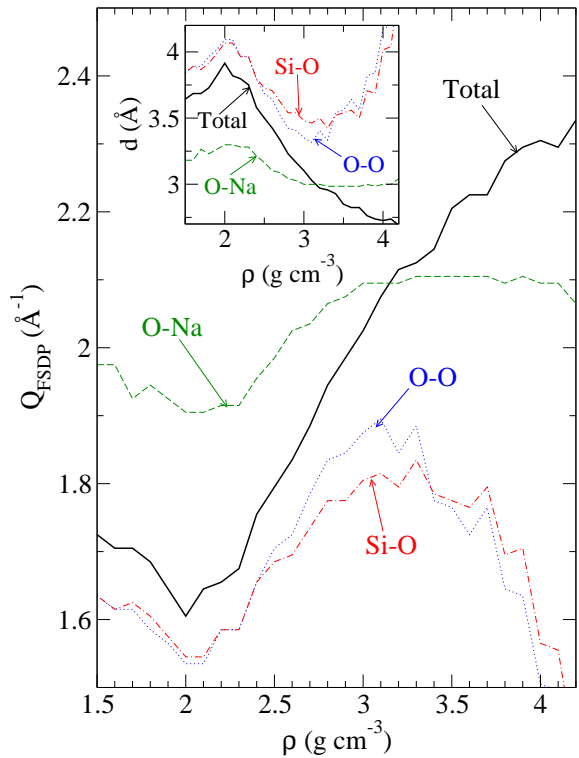,width=\linewidth}
\end{center}
\caption{\label{qfsdp} (Color online) FSDP position of the total Neutron structure factor and positions of each relevant partial structure factors FSDP. The insert shows the associated characteristic distance $d=2\pi/Q_{\text{FSDP}}$ of the total and partial structure factors.}
\end{figure}

\begin{figure}
\begin{center}
\epsfig{figure=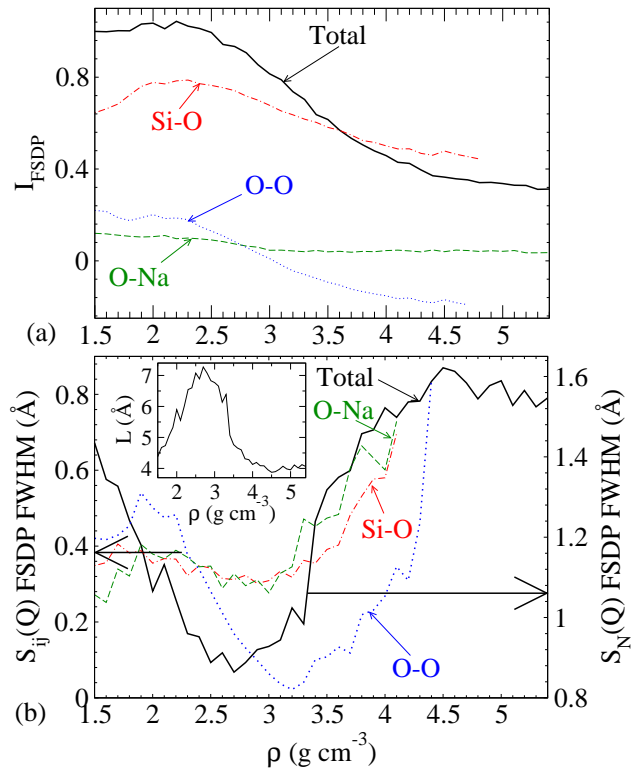,width=\linewidth}
\end{center}
\caption{\label{IFWMHfsdp} (Color online) (a) Intensity of the FSDP of the total and partial structure factors. (b) FSDP FWHM of the total and partial structure factors. The insert shows the correlation length $L=2\pi/\text{FWHM}$.}
\end{figure}

The FSDPs we obtained from simulations were further studied by fitting them with Lorentzian functions (examples of fitted functions can be seen on Fig. \ref{S}). This choice is supported by the fact that the experimental results in neutron scattering factor of silica can be better fitted with a Lorentzian function than with a Gaussian one\cite{wright_structure_1991}. It should be noted that the fit has been done on the low $Q$ part of the FSDP to avoid the contribution of the following peaks. This allows to track precisely intensity, position and full-width at half maximum (FWHM) with density. Fig. \ref{qfsdp} and \ref{IFWMHfsdp} show the position $Q_{\text{FSDP}}$, the intensity $I_{\text{FSDP}}$ and the FWHM of the FSDP. The computed FSDP position at room density ($1.85\mathring{\text{A}}$$^{-1}$) is found to be in very good agreement with the one obtained from experiment ($1.83\mathring{\text{A}}$$^{-1}$) \cite{gaskell_medium-range_1996} and with the one from  the MD work by Corrales et al. ($1.77\mathring{\text{A}}$$^{-1}$)\cite{du_compositional_2006}. Except at very low density ($\varrho <$ 2.5 g/cm$^3$, negative pressure domain), the FSDP position increases with density while its intensity decreases. This trend is consistent with X-ray diffraction results from Benmore in densified silica.\cite{benmore_structural_2010} Interestingly, the FHWM of the FSDP exhibits a density window between 2.3 and 3.3 g/cm$^{3}$ with a minimum found at $\varrho$ = 2.7 g/cm$^3$.

Coming back to the real space correlations, the FSDP peak position $Q_{\text{FSDP}}$ is usually related to a characteristic repeat distance $d=2\pi/Q_{\text{FSDP}}$ and the FWHM to a correlation length $L=2\pi/\text{FWHM}$, sometimes also called 'coherence length', due to atomic density fluctuations \cite{gottlicher_compositional_1996, zotov_effect_1996}. The effect of irradiation \cite{susman_structural_1990, wright_neutron_1991}, water content \cite{zotov_effect_1996, zotov_effect_1992} and alkali content \cite{du_compositional_2006} on the FSDP have been studied, leading to the idea that a depolymerization of the network (a decrease of the atomic order) is associated to a decrease of the intensity of the FSDP and a decrease in the characteristic distance $d$. A global understanding of the correlation length $L$ is lacking since it has been found to decrease with increasing potassium amount in silica network, to increase with increasing lithium amount and to show a maximum in sodium silicate when $x=0.20$ \cite{du_compositional_2006}. It seems therefore highly system dependent.

The inserts of Fig. \ref{qfsdp} and \ref{IFWMHfsdp}b show the computed characteristic distance $d$ and characteristic correlation length $L$ as a function of density. It can be observed that the characteristic distance $d$ decreases with density, which suggests a decrease of the medium range order (MRO). On the contrary, the correlation length $L$ does not follow a general behavior with density since it shows a density window between 2.3 and 3.3 g/cm$^{3}$ with a maximum at 2.7 g/cm$^{3}$. We notice that the density of the maximum of $L$ corresponds to the density of the beginning of the growth of the Si$^{\text{V}}$ fraction (see Fig. \ref{n}a).

\subsubsection{Contributions to the FSDP}

Even though it can be noticed from Fig. \ref{Sij} that all partial structure factors show a FSDP, they do not contribute to the FSDP of the total structure factor at the same level. To understand the behavior of the FSDP, the position, intensity and FWHM of the FSDPs of each partial structure factor $S_{ij}(Q)$ have been computed.

Fig. \ref{qfsdp} and \ref{IFWMHfsdp}a show the position and the intensity of the FSDPs according to density. At low density ($\varrho <$ 2.7g/cm$^3$), the main contribution to the total FSDP clearly comes from the Si-O FSDP, since their position and intensity are similar and show the same trend. However, at larger densities, the partial FSDPs positions show a maximum (around $\varrho$ = 3.1g/cm$^3$) whereas the total FSDP continuously increases. These maximums correspond to minimums of the characteristic repeat distance $d$ and we notice that they occur at the density at which the Si$^{\text{VI}}$ fraction starts to grow. This shows that the total FSDP is not a simple superposition of the partial FSDP. The shift of the total FSDP to higher $Q$ at high density can mainly be explained by the increased contribution of the main peak of the $S_{\text{O-O}}$ (at 3$\mathring{\text{A}}$$^{-1}$) whose intensity grows with the density.

Even though the link between the total and the partial FSDPs is not simple, it is interesting to notice that each partial FSDPs show a minimum of their FWHMs according to the density (see Fig. \ref{IFWMHfsdp}b). Si-O and O-Na partial FSDPs FWHM reach their minimums around 2.7g/cm$^3$, corresponding to the minimum of the total FSDP FWHM. The O-O partial FSDP shows the sharpest minimum of its FWHM around 3.2g/cm$^3$ (i.e. at larger density than for the total FSDP) but does not contribute a lot to the total FSDP due to its low intensity (see Fig. \ref{IFWMHfsdp}a).

\subsection{Vibrational properties}

The nature of the vibrational excitations of silicate glasses has so far remained a challenging issue. As contrary to crystals, the lack of long-range structural order in amorphous solids strongly affects their vibrational dynamics. The appearance of an excess of vibrational modes over the Debye level at terahertz frequencies, the so-called Boson peak (BP), is one of the special features exhibited by glasses.

\subsubsection{Vibrational density of states}

\begin{figure}
\begin{center}
\epsfig{figure=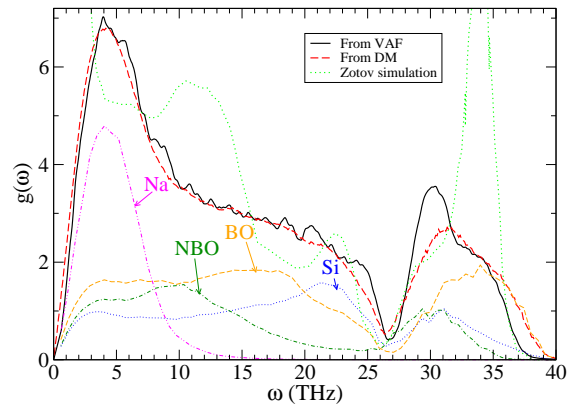,height=\linewidth, angle=-90}
\end{center}
\caption{\label{vdos0} (Color online) Vibrational density of states at room pressure computed from the Fourier transform of the velocity autocorrelation function (VAF) and from the diagonalization of the dynamical matrix (DM). The results are compared with the one from the simulation of Zotov et al.\cite{n_effects_2001} using a different potential. The partial VDOS for Si, BO, NBO and Na are also shown.}
\end{figure}

\begin{figure}
\begin{center}
\epsfig{figure=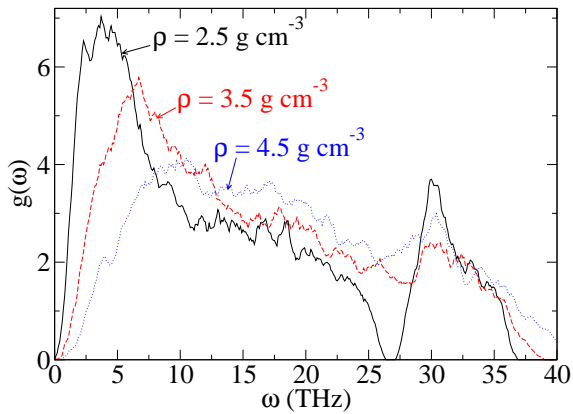,height=\linewidth, angle=-90}
\end{center}
\caption{\label{vdos} (Color online) Vibrational density of states, computed from the VAF, for different selected densities.}
\end{figure}

The vibrational density of states (VDOS) $g( \omega )$ can be computed in two different ways. Starting from a relaxed glass (via energy minimization or cooling to 0K), one can compute the dynamical matrix (DM) by evaluating the second derivative of the total energy with respect to small atomic displacements\cite{mihailova_vibrational_1994}. The diagonalization of the DM provides the eigenvalues, i.e. the frequency of each normal vibrational mode. Another way is to compute the Fourier transform of the velocity autocorrelation function (VAF) :

\begin{equation}
 g( \omega ) = \frac{1}{N k_B T} \sum \limits_{j=1}^N m_j \int_{-\infty}^{\infty} <\textbf{v}_j(t) \textbf{v}_j(0)> \text{exp}(\text{i}\omega t)\, \mathrm dt
\end{equation} where $N$ is the number of atoms, $m_j$ is the mass of an atom $j$,  $\omega$ is the frequency and $\textbf{v}_j(t)$ is the velocity of an atom $j$. It has been reported that both methods lead to quite similar VDOS in silica\cite{oligschleger_dynamics_1999}. Although the DM approach is far more expensive computationally than the VAF one, it should be noted that, looking at the eigenvectors $\textbf{e}_i$ associated to each eigenvalue frequency $\omega_i$, one can get the details of the nature of each normal mode. One can, for example, compute the partial VDOS $g_{\alpha}( \omega )$ ($\alpha$ = Si, BO, NBO, Na) for each atom defined as :

\begin{equation}
  \label{partialvdos}
  g_{\alpha}( \omega_i ) = g( \omega_i ) \sum \limits_{j \in \alpha} |\textbf{e}_j (\omega_i)|^2
\end{equation} where $\textbf{e}_j (\omega_i)$ are the 3-component real space eigenvectors associated to the atoms $\alpha$.

Fig. \ref{vdos0} shows the VDOS, scaled to one, computed using the two previously described methods. We observe a fair agreement between both methods, especially at low and intermediate frequency. The difference at high frequency can be explained by the harmonic assumption on which the DM method relies. The results are compared with the one from the simulation of Zotov et al.\cite{n_effects_2001}. We observe that the agreement is very poor even if some trends are similar (sharp peak at high frequency and appearance of a new peak at low frequency increasing with respect to the amount of sodium). We note that the simulation from Zotov et al. uses a different potential (from Vessal) involving both 2- and 3-body terms and that the system is quite smaller (1080 atoms, as compared to 3000 in the present simulation). Unfortunately, to our knowledge, no experimental VDOS is currently available for this composition.

Using Eq. \ref{partialvdos}, the partial VDOS have been computed and are shown on Fig. \ref{vdos0}. Note that O atoms have been split into BOs and NBOs. Relying on the vibrational analysis of silica \cite{oligschleger_dynamics_1999, taraskin_nature_1997}, one can interpret some features of the present VDOS. Si atoms contribute the most at high frequency (27-37 THz, symmetric and anti-symmetric stretching modes) and at intermediate frequency (22 THz, O-Si-O bending mode). BO atoms predominant contributions occur at high frequency (29-37 THz, symmetric stretching modes) and at low and intermediate frequency (0-26 THz, Si-BO-Si bending mode and symmetric stretching mode). The contribution of NBOs differs from the one of the BOs because of the decreased number of O-Si stretching modes and of the softening of the Si-NBO-Na bending mode as compared to the Si-BO-Si mode. Most of the low frequency contribution comes from Na atoms (0-10 THz, Na-NBO low-energy stretching modes).

Fig. \ref{vdos} shows the vibrational density of states, scaled to one, for different increasing densities. We observe some trends which are similar to the ones observed in densified silica \cite{f._vibrational_2011} : decrease of the number of low-frequency modes, disappearance of the gap between intermediate and high frequency modes around 27 THz and broadening of the high-frequency peak. The low-frequency region, whose contribution mainly comes from Na atoms, is the most affected part of the VDOS, suggesting that the Na vibrational modes are strongly modified during densification.

As density increases, the low-frequency peak coming from Na-O bounds decreases in intensity and is shifted to higher frequencies. The high frequency peak coming from Si-O bounds becomes broader but does not show any significant frequency shift.

\subsubsection{Boson peak}

\begin{figure}
\begin{center}
\epsfig{figure=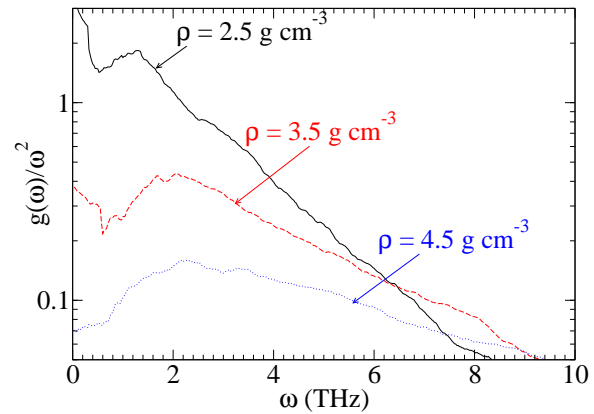,height=\linewidth, angle=-90}
\end{center}
\caption{\label{BP} (Color online) Boson peak visualization for different selected densities.}
\end{figure} 

\begin{figure}
\begin{center}
\epsfig{figure=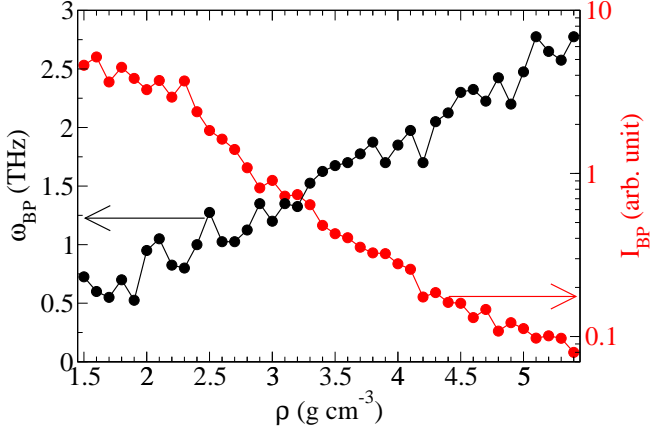,height=\linewidth, angle=-90}
\end{center}
\caption{\label{BP2} (Color online) Boson peak position (left axis) and intensity (right axis) with respect to density.}
\end{figure}

The origin of the BP in silica, still controversial, has been associated to the existence of local modes involving rocking motions of distorted SiO$_4$ tetrahedrons \cite{buchenau_low-frequency_1986, courtens_vibrational_2001, guillot_boson_1997}. The BP can be observed by looking at the excess VDOS over the Debye law $g(\omega) - g_{\text{D}}(\omega)$ or at the quantity $g(\omega)/\omega^2$. The latter quantity can be identified with the one-phonon scattering cross section as measured in neutron scattering experiments \cite{guillot_boson_1997, carpenter_correlated_1985} and is shown on Fig. \ref{BP} for three selected densities. A pronounced peak can be observed at each density, even if its intensity decreases with density. At room pressure, the BP is found to be located at $\omega_{\text{BP}}$=1.3 THz, which is lower than the value found experimentally (Raman) of 1.95 THz (65 cm$^{-1}$) \cite{mcintosh_boson_1997}.

The BP properties have been further analyzed by computing its position $\omega_{\text{BP}}$ and its intensity $I_{\text{BP}}$, quantities that are displayed on Fig. \ref{BP2}. It can been observed that $I_{\text{BP}}$ decreases with density, while $\omega_{\text{BP}}$ increases with density. Both of these two trends (decrease of the intensity and increase of the frequency) have been observed experimentally in many system, such as in pure silica \cite{zanatta_elastic_2010}, in lithium silicate glass \cite{kitamura_high_2000}, in a Na$_2$FeSi$_3$O$_8$ glass \cite{monaco_effect_2006} as well as in different polymers \cite{hong_pressure_2008}.

\section{Liquid}
\subsection{Thermodynamics}

\begin{figure}
\begin{center}
\epsfig{figure=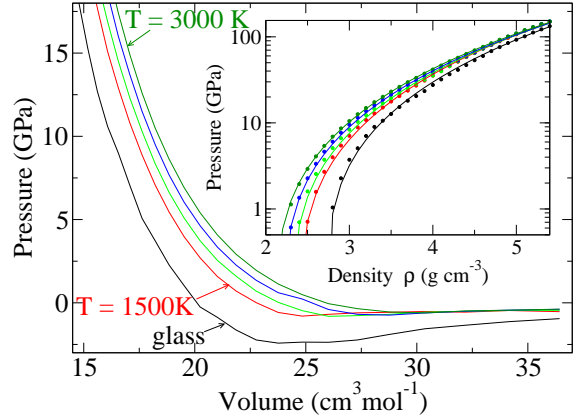,height=\linewidth, angle=-90}
\end{center}
\caption{\label{EOS} (Color online) Isotherms for glass and liquid NS2. The curves are separated by 500K each. The inset shows the corresponding data in ($\varrho$, P) together with the BM fits (solid lines).}
\end{figure}

\begin{figure}
\begin{center}
\epsfig{figure=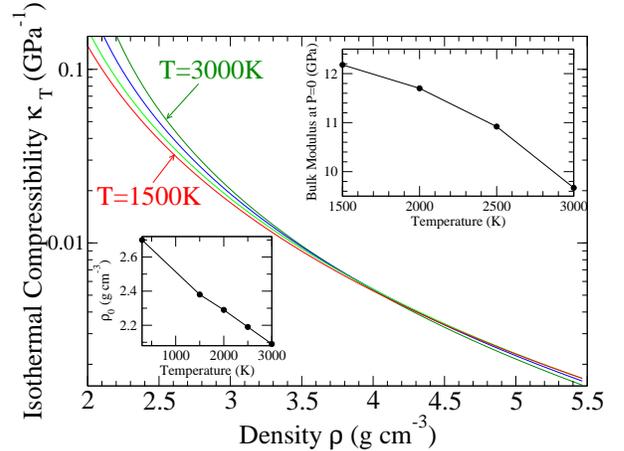,height=\linewidth, angle=-90}
\end{center}
\caption{\label{khi} (Color online) Isothermal compressibility $\kappa_T$ with respect to density in liquid NS2 for various temperatures ranging from 3000 to 1500K separated by 500K each. The curves are computed using the BM EOS. The inset shows the room pressure density $\varrho_0$ change with temperature.}
\end{figure}

To evaluate the equation of state (EOS) of the liquid, many thermodynamics points have been computed (T, $\varrho$, P). The following range has been studied : $1.5 \leqslant \varrho \leqslant 5.5$ g/cm$^3$ and $1500 \leqslant $ T $ \leqslant 3000$ K, which correspond to the following pressure range : $-2.23 \leqslant $ P $ \leqslant 150$ GPa. In contrast with previous works on molecular fluids \cite{guissani_computer_1993, ree_analytic_1980} and silica, where the data were fitted using a Van der Waals type EOS, the data of the current simulations were fitted with a Birch-Murnaghan equation of state (BM EOS) that has a simpler form \cite{birch_elasticity_1952, s._k._saxena_thermodynamics_1993}. It has revealed to give reasonable fits in the case of a liquid densified germania \cite{micoulaut_simulated_2006} and is widely used in geophysical studies (see for example \cite{agee_crystal-liquid_1998}).

Fig. \ref{EOS} shows the isotherms of the glass and the liquid in the (P, V) representation. The data have been fitted far from the critical region with the BM EOS, that has the following form :

\begin{equation}
 P=\frac{3}{2} K ( ( \frac{\varrho}{\varrho_0} ) ^{7/3}-[ \frac{\varrho}{\varrho_0}]^{5/3}) ( 1-\frac{3}{4} ( 4-K_1 ) ([ \frac{\varrho}{\varrho_0} ] ^{2/3}-1 ) )
\end{equation} where $K$ is the bulk modulus at P=0, $K_1=dK/dP$ at P = 0 and $\varrho_0$ is zero-pressure density of the liquid. The fit can be made with two parameters only ($K$ and $K_1$) since $\varrho_0$ can be accessed from the isothermal data displayed on Fig. \ref{EOS}. It can be seen that the data are very well fitted by the BM EOS along all the density and temperature range.

In addition, the BM EOS allows to have access to the bulk modulus $K$ at P=0 and to the isothermal compressibility $\kappa_T=\varrho^{-1}(\partial \varrho / \partial P)_T$ according to the density (plotted on Fig. \ref{khi}). The observed behavior, enhanced compressibility with falling density, is realistic, as well as the decrease of the bulk modulus at P = 0 with respect to the temperature (see the insert of Fig. \ref{khi}). The results are in good agreement with the only experimental data on liquid NS2 we are aware of ($K$ = 13.4 GPa and $\kappa_T$ = 0.075 GPa$^{-1}$ at T=1500 K and P = 0)\cite{webb_compressibility_1996}.

\begin{figure}[t!]
\begin{center}
\epsfig{figure=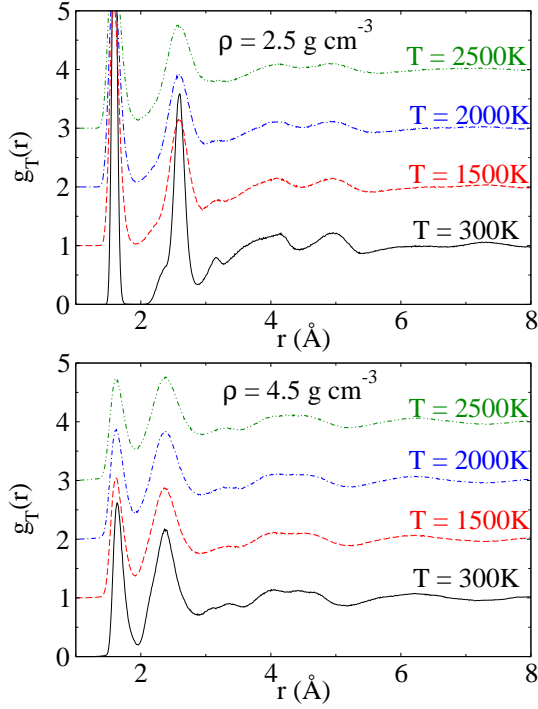,width=\linewidth}
\end{center}
\caption{\label{g2} (Color online) Radial distribution function for increasing temperatures.}
\end{figure} 

\begin{figure}[t!]
\begin{center}
\epsfig{figure=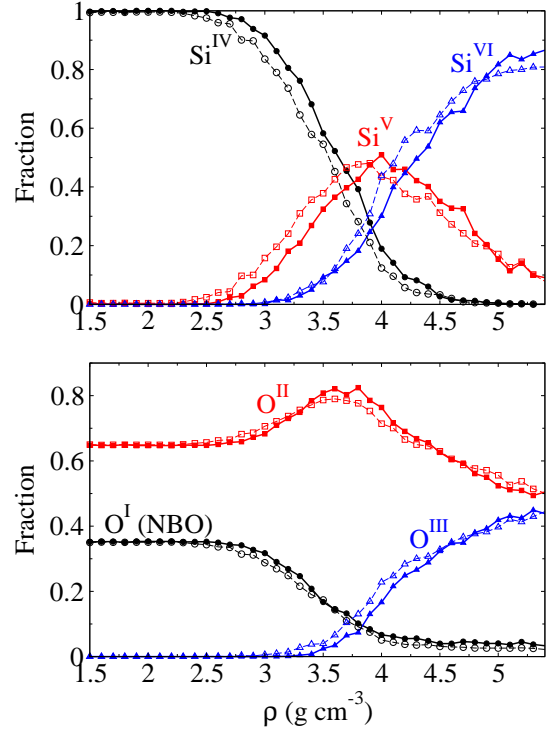,width=\linewidth}
\end{center}
\caption{\label{n2} (Color online) Distribution of IV, V and VI-fold coordinated silicon atoms (a) and of I, II and III-fold coordinated oxygen atoms (b) with respect to density both at 300K (filled symbols) and 2000K (open symbols). Sodium atoms are not taken into account in the enumeration of the neighbors, so that O$^{\text{I}}$ refer to NBOs.}
\end{figure}

\subsection{Structure}

Fig. \ref{g2} shows the total correlation function $g_{\text{T}}(r)$ for different increasing temperatures, both at low and high density. It can be observed that density seems to have a more critical influence on structure than temperature. Indeed, the positions of the peaks do not show any significant shift as temperature increases. The only behavior that can be seen is a broadening of all peaks as temperature increases, which can be explained by the increasing disorder due to the increasing thermal energy. The same trends are observed in pure silica \cite{susman_temperature_1991}.

Eventually, the influence of the temperature on the coordination numbers has been checked. The populations of the different Si and O species according to the density are plotted on Fig. \ref{n2} both at T = 300K and 2000K. The transition between Si$^{\text{IV}}$ into Si$^{\text{V}}$ and Si$^{\text{VI}}$ is still clearly observed, although the transition occurs at a lower density than in the glass (density shift of approximately 0.2 g/cm$^3$). The same shift can be observed for the O species.

\section{Conclusion}

Our purpose in the present paper has been to provide a systematic and extensive study  of the properties of densified glassy and liquid NS2 sodium silicate.

While bond distances remains nearly unchanged, pressure has a strong effect on angles and coordination numbers. A transition from tetrahedral to octahedral silicon environment is found. The fraction of NBOs decreases and O$^{\text{III}}$ tricluster are observed at high density. The usual vibrational behavior is observed, i.e. the decrease of the amount of low frequency modes, the increase of the frequency of the Boson peak and the decrease of its intensity under pressure. Expected anomalous effects are found in the medium range order (increase of the position of the FSDP and decrease of its intensity under pressure), but, more surprisingly, we observe a minimum of the FWHM of the FSDP according to the density. Temperature is found to have only small effects on structure and the Birch-Murnaghan equation of state allows to reproduce the densification of the liquid at each temperature.

Finally, it is worth mentioning that, as ambient pressure, it is well-known that changes in composition of the glass (and especially in the amount of sodium atoms) induce changes of the degree of polymerization of the glass. These competitive effects (depolymerization by sodium atoms and repolymerization by the pressure) should be addressed in the future for a better understanding of the glass network properties.

\begin{acknowledgments}
Warm thanks are due to M. Micoulaut for suggesting this study and providing advice at its various stages, to J. C. Mauro for his help to compute the Dynamical Matrix and to G. Mountjoy for a very stimulating discussion.
\end{acknowledgments}

\bibliography{silicate}


\end{document}